\documentclass[twocolumn,showpacs,amsmath,amssymb]{revtex4}
\usepackage[dvips]{graphicx}
\usepackage{dcolumn}
\usepackage{bm}

\begin{document}

\title{Continuous loading of an atom beam
 into an optical lattice.}

\author{Vladyslav V. Ivanov}
\email{vladivanov78@gmail.com}
\affiliation{24 Princeton street, North Chelmsford, MA, 01863, USA}

\date{\today}

\begin{abstract}

I propose a method of deceleration and continuous loading of an atom beam into a far-off-resonance optical lattice.
The loading of moving atoms into a conservative far-off-resonance potential requires the
removal of the atom's excess kinetic energy. Here this is achieved by
the Sisyphus cooling method, where a differential lattice-induced ac Stark shift is utilized.
The proposed method is described for the case of ytterbium atoms.
Numerical simulations demonstrate the possibility of reaching cold and dense samples in a continuous manner on the example of ytterbium atoms.

\end{abstract}



\maketitle

\section{Introduction}


Matter-wave interferometry with ultra-cold atoms holds great promises for precision measurements. Pioneering experiments have demonstrated the possibilities of the measurements of photon recoil and Earths gravity acceleration \cite{Weiss93,Peters1999,Gupta2002,Ferrari2006}. More recently, very accurate measurements of the fine structure constant have been performed in group of Francois Biraben \cite{Bouchendira11}. Atom interferometry was used for measurements of
the gravitational constant \cite{Fixler07,Lamporesi08}, the least accurately known fundamental constant.
 Although matter-wave interferometry greatly benefits from short de Broglie wavelength of laser cooled atoms, it is still far behind in terms of flux, i.e. the number of available particles per time-interval, when compared to more the common optical interferometers. Obtaining a bright source of ultra-cold atoms can be crucial for further advances in matter-wave interferometry.

Traditionally, atoms used in matter-wave interferometry experiments are, first, loaded into magneto optical traps (MOT), followed by an intermediate polarization gradient cooling stage or a compressed MOT stage. Then these atoms either are immediately used for measurements \cite{Peters1999,Bouchendira11,Fixler07} or loaded into conservative traps for further manipulation \cite{Gupta2002,Ferrari2006}.  However, such methods imply a time sequence of cooling steps, which restricts the loading rate of atoms.
Continuous production of ultra-cold atoms can increase the available atomic flux for matter-wave interferometry experiments.
 Furthermore, it can
potentially offer  the possibility of preparing Bose-Einstein condensates (BEC) continuously with a substantially increased flux.
A spectacular application of this would be the realization of an atom laser, the matter wave
analogue to an optical cw laser.
A continuous atom laser would provide an extremely bright and coherent source of matter that promise an significant improvements of precision measurements and might open novel ways for fundamental tests of quantum mechanics \cite{Robins08}.

In this paper I propose the method of a dissipative deceleration and continuous loading of an atomic beam into a standing wave potential of optical lattices.
The method includes a Sisyphus cooling mechanism which allows the removal of excessive kinetic energy. The proposed method exploits the differential ac Stark shift
induced by the optical lattice potential, similarly to \cite{Ivanov2011}. Atoms are  pumped into the excited state using an additional pumping beam that is tuned near a resonance  at
the node of the lattice potential.  In the excited state, atoms dissipate their kinetic energy by climbing the "hills" of the lattice potential. Eventually, atoms decay into the ground state and this cycle repeats until the atoms have spent all their energy and are trapped in anti-nodes of the optical lattice.
Atoms trapped in the conservative optical lattice potential are essentially sitting in the "darkness" due the large ac Stark shift.
The presented Sisyphus cooling mechanism is different from other Sisyphus cooling schemes demonstrated previously \cite{Pritchard1983,Lett88,Newbury1995,Ovchinnikov1997,Miller2002,Ferrari2001} in two regards: 1. Atoms do not need to be pre-cooled or trapped, 2. Atoms cooled below a certain energy do not scatter cooling photons and thus they are essentially decoupled from cooling cycle.
The described scheme can be modified by introducing a small frequency difference for the optical lattice beams, creating an "atomic
conveyor belt" \cite{Schrader2001,Schmid2006}, which allows continuous production of cold and dense atom samples.

The proposal is described for the case of ytterbium atoms, although an application of the proposed method on other atomic species is possible.
Ultra-cold samples of Yb atoms are
of great interest in the context of precision measurements including optical clocks \cite{Ybclock} and matter-wave interferometry \cite{Alan2011,Ivanov2012}.

The method, I describe, is similar to the scheme proposed by A. Aghajani-Talesh \cite{Aghajani09}, where the dissipation of atom energy is achieved by pumping atoms into different magnetic states at the top of the magnetic barrier.
However, the scheme was proven to be effective in a recent experiment for loading atoms into an optical dipole trap (ODT)  \cite{Falkenau2011}. This scheme is limited to atomic species with magnetic sub-levels in the ground state. Moreover, although an injection of an atomic beam into an ODT was demonstrated, uncoupling of the atoms cooled in the ODT in a continuous way still has not been proposed.

The rest of this paper is organized as follows. In Sec. II, I discuss the general requirements for the proposed loading method. I perform a simple calculation for the efficiency of the loading process and discuss its limitations. In Sec. III
I develop a numerical model that addresses the stochastic nature of the process and computes expected atom dynamics.
In Sec. IV, I discuss reachable loading rates and density of atomic samples in an experiment with realistic parameters. I draw my concluding remarks in Sec. V.

\section{The Deceleration Scheme}

\begin{figure}[ht]
\includegraphics[width=78mm]{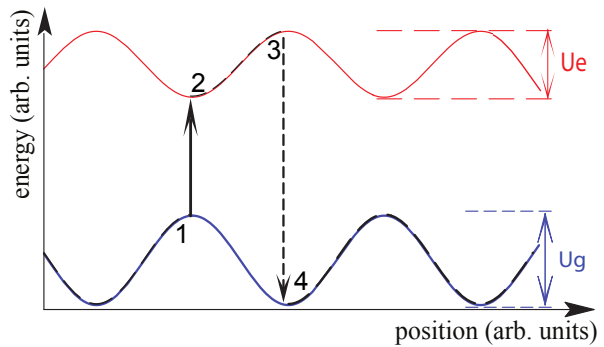}\rule{5mm}{0pt}
\caption{\label{fig:Scheme1} (color online) Continuous loading scheme for a two-state cooled species in optical lattices. An additional pumping beam,  resonant with the transition, is also directed at the trapping region. The solid upper red (lower blue) curve shows the spatially-varying lattice potential, where $U_e(U_g)$ is the lattice potential amplitude for the excited (ground) state.
The pumping laser is tuned near the resonance for an atom at the zero of the trapping potential, i. e. nodes of the optical lattice. The cooling proceeds via the following cycle: 1. a moving atom absorbs a photon at a node of the lattice potential; 2. it climbs a repulsive potential (upper red curve); 3. it spontaneously decays into the ground state  (lower blue curve); 4. as the atom continues to move in the lattice potential, it absorbs another photon near the node of the lattice and the cycle repeats.
Trajectory of the atom is shown as a dashed black curve.
 }
\end{figure}

The basic idea of the deceleration scheme is shown in Fig.~\ref{fig:Scheme1}. An atomic beam is moving towards a trapping region where the optical lattice beams and pumping beams intersect. The pumping beams are tuned near the resonance for the atoms at the nodes of the optical lattice and is perpendicular to  the atomic beam. In the transverse direction the pumping beams assumed to have diameters substantially larger that the optical lattice, the diameters of pumping beams in longitudinal direction essentially define the deceleration region.
The optical lattice is aligned along the atom beam or crosses it at a very shallow angle. The optical lattice potential, which is attractive for the atom in the ground state but repulsive in the excited state, is playing a double role: 1. it provides potential hills on which atoms spend their energy 2. it provides trapping potential for atoms that have already dissipated their kinetic energy. When an atom scatters one pumping photon, it leads to a loss of atomic energy roughly equal to lattice trap depth. This energy loss can be much larger compared to the energy loss during Doppler cooling, which is proportional to a photon recoil momentum. After the atom energy drops below the lattice trap depth, the atom is trapped in a lattice site and does not scatter any more cooling photons due to the ac Stark shift.
The proposed method critically relies on the spatially selective pumping of atoms into the excited state, which is achieved using the ac Stark
shift provided by the optical lattice potential.

An important requirement of this method is that the ac Stark shift of the excited state should be larger than the ground, i.e. the optical lattice
for the atoms in the excited state has to be either repulsive or if it is attractive it must have smaller ac Stark shift than for the ground state.
The linewidth of the pumping transition is another crucial parameter.  Relatively  narrow transitions with a linewidth below 1 MHz are preferable because they allow pumping of the atoms precisely at the nodes of an optical lattice.
As it is shown below, another advantage of a narrow transition is a longer decay time into the ground state, which makes the deceleration method efficient for lower velocities (see Eq.4). However, broad transitions can provide high scattering rate, which would increase probability of pumping into the excited state, and thus increase  capturing velocity (Eq.3). The linewidth of the pumping transition is a trade-off between higher captue velocity
and ability to decelerate atoms to lower velocities.
For realistic experimental parameters the described method will work efficiently for pumping transition in the range between $ 2 \pi \times 10$ kHz and $ 2 \pi \times 1$ MHz.

To estimate a maximum capturing velocity and number of scattered photons I first consider the 1D case.
Let's assume photon absorption occurs strictly at the nodes of the optical lattice.
If the atom velocity is high, the atom can propagate through the lattice potential for distances substantially larger then the lattice period during the typical time scale of spontaneous decay into the ground state. In this case the position of where the atom decays into the ground state is not correlated with the position of the lattice potential minima or maxima.
In this approximation the atom loses the average energy $U_{d}=(U_{e}-U_{g})/2$ per cycle. In order to decelerate an atom with the initial kinetic energy $E_{0}$,
the atom should scatter $ \sim  E_{0}/U_{d}$ photons.
For a deceleration region with size $l_{d}$, the number of trapping lattice sites $N_{s}$ is about $2l/\lambda$, where the $\lambda$ is the lattice wavelength.
In the deep lattice approximation the probability to absorb a cooling photon can be estimated as
\begin{equation}
p=\frac{\gamma \lambda}{2 v_{0}} \sqrt{\frac{\hbar \gamma}{U_{d}}},
 \label{eq:absprobability}
\end{equation}

where $v_{0}$ is the atom velocity and $\gamma$ is the linewidth of the transition.
The atom moving through the cooling region will dissipate the energy $E_{d} =  p \times N_{s}\times U_{d}$, on the order of

\begin{equation}
E_{d}\simeq \frac{\gamma \lambda}{2 v_{0}}\sqrt{\frac{\hbar \gamma}{U_{d}}}\frac{2l_{d} }{\lambda}U_{d}.
 \label{eq:dissipatedenergy}
\end{equation}

After atom energy drops below $U_{g}$, the atom stops propagation and is trapped in the lattice potential.
The initial kinetic energy dissipated in such a system is

\begin{equation}
v_{0}^{6} \simeq 4 \gamma^{2} l_{d}^{2} \times \hbar \gamma \times U_{d}/m^{2}.
 \label{eq:maxtemperature}
\end{equation}

The maximum capture velocity does not depend on the lattice wavelength.
It is worth noting that this result is valid only the deep optical lattice, i.e. $\hbar \Gamma \ll U_{d}$.
For the case of $^{174}$Yb, $U_{d}=1$ mK, and $l_{d}$=5 cm, $v_{0} \simeq 8$ m/sec, atoms with this initial velocity or lower will be decelerated and trapped. This velocity is comparable to the capture velocity reached in conventional Yb MOT \cite{Kuwamoto1999}.
The described method becomes less efficient for lower velocities. When the assumption that atoms in the excited state propagate through the lattice potential for distances substantially larger then the lattice period fails, the dissipative process becomes inefficient.
The expected energy $\varepsilon$  the atoms lose after scattering a photon at the node of the lattice potential is (assuming that change of atom's velocity is insignificant)

\begin{equation}
\varepsilon=\int_{0}^{\infty}\gamma \textbf{e}^{-\gamma t}(U_{p}-U_{s})\sin^{2}(k v t)dt=  \\
\frac{8 \pi^{2} v^{2}}{16 \pi^{2} v^{2}+\gamma^{2}\lambda^{2}},
 \label{eq:maxtemperature}
\end{equation}
where $k=2 \pi / \lambda$.
At the velocities of $v\simeq \gamma \lambda/ (4 \pi)=0.1$ m/sec the efficiency starts to drop.

\section{Numerical Model}

Absorption and spontaneous emission of photons are
random processes. To investigate the stochastic nature of atomic trajectories a semiclassical Monte Carlo model was developed.
The numerical model is applied to the dynamics of a single atom moving through the interaction region and it assumes that for a small enough time increment, changes in the relevant parameters such as atom position and velocity are negligible. The atom motion is broken into a set of discrete time-steps $\Delta t$, and at each step the probability of the atom to make a transition (either absorption or spontaneous emission) is calculated. I calculate the new position and velocity assuming that the acceleration is constant during $\Delta t$. If a transition occurs during a particular step, the atom starts the next step in the other state and experiences the different trapping potential. $\Delta t$ is chosen to be 20 ns or less, which is much smaller than all other time scales for all the calculations in this paper.
For an atom in the ground state,  the absorption of a cooling photon is a random process that happens with
probability given by the right-hand-side of Eq. \ref{eq:probabs}.
\begin{equation}
p(\textbf{r})=\frac{\gamma}{2} \frac{s \times \Delta t}{1+s+\left(\frac{\delta-\delta_{l}
(\textbf{r})}{\gamma/2}\right)^{2}}.
\label{eq:probabs}
\end{equation}
where $s=I/I_{\rm sat}$ is the saturation parameter (intensity in units of the saturation intensity) of the pumping laser, and $\delta$ is its detuning from the transition frequency of the free atom. $\delta_{l}(\textbf{r})$ is the lattice induced differential ac Stark shift between the excited and ground states.

For the following examples I assume that the optical lattice was created by a retro-reflected beam at $\lambda$=1064 nm. For this wavelength, the polarizability of $^{174}$Yb in $^{1}S_{0}$  state is $\alpha_{g}=372.2\times10^{-39}$ $Cm^{2}/V$, and in  $^{3}P_{1}$ is  $\alpha_{e}=-82.3\times10^{-39}$ $Cm^{2}/V$. The trap depth of the optical lattice with beam waist of 50 $\mu$m for atoms in the ground state is about 550 $\mu$K when the optical power is P=20 W.
The following numerical simulations are averaged over the individual atom trajectories and atom-atom interaction is ignored.

\subsection{1D case}

The basic properties of the proposed scheme are demonstrated using the 1 dimensional example. Since absorption and spontaneous emission of photons occur randomly, each individual atom's trajectory is different and has abrupt changes when photon scattering happens. However, after averaging over 100 or more cases, curves look smooth and some general consistency appears.

In Fig. 2 the dependence of the atom velocities versus time is shown.
The lattice potentials for ground (excited) states are written as: $U_{g(e)}(z)=U_{0g(e)}\sin^{2}(k z)$.
The atoms are decelerated until their kinetic energy falls below the lattice potential trap depth of 549 $\mu$K (20W optical power).  At this point they are trapped in the lattice sites. It takes longer to decelerate atoms with a higher initial velocity.

The probability of absorbing a pumping photon strongly oscillates depending on the relative position of the atom with respect to a node of the optical lattice. However after many averages this effect slowly washes out. I present the dependence of scattering rate versus time in the Fig.3 for the same conditions as for Fig.2, averaged over the large number of
 atom trajectories and for time periods of 20 $\mu$s.  Each point in Fig. 3. represent not only the average of 500 iterations, but also an average over 100 20 ns time-steps.  Atoms moving through the optical lattice scatter pumping photons with a rate of about 40 photons per millisecond.  After being decelerated and trapped, scattering rates drop below 2-3 photons per millisecond.

The deceleration length  $l_{d}$, i.e. the distance an atom must propagate before it is trapped, depends on various parameters such as optical lattice depth, linewidth of the pumping transition $e.t.c$. The dependence of the deceleration length on initial atom velocity is shown in Fig.4. To compute Fig. 4 about 15 average trajectories were simulated for different initial velocities, each of which is an average of 20 single atom trajectories. Deceleration lengths
 were extracted and then interpolated by polynom of 7th order, however approximately $l_{d}$ scales $v_{0}^{3}$ as expected from Eq.3.

\begin{figure}[ht]
\includegraphics[width=74mm]{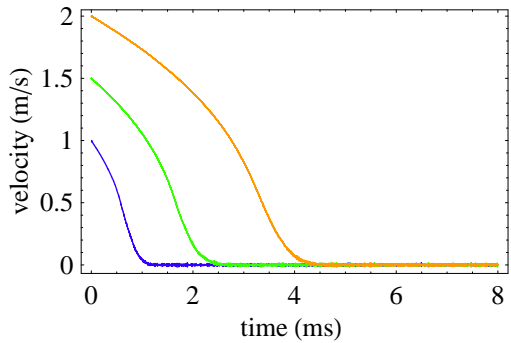}\rule{5mm}{0pt}
\caption{\label{fig:VzvsT} (color online) Averaged velocity of Yb atom versus time. Here $\Delta t = 20$ ns. The lattice potentials is created by a 20 W retro-reflected beam focused down to a 50 $\mu$m waist at $\lambda$=1064 nm. The pumping beam is tuned near resonance with $\delta=- \gamma/2$ and s=5. The three curves correspond to initial velocities of 1, 1.5, 2 m/s (blue, green and orange curves, respectively). }
\end{figure}

\begin{figure}[ht]
\includegraphics[width=78mm]{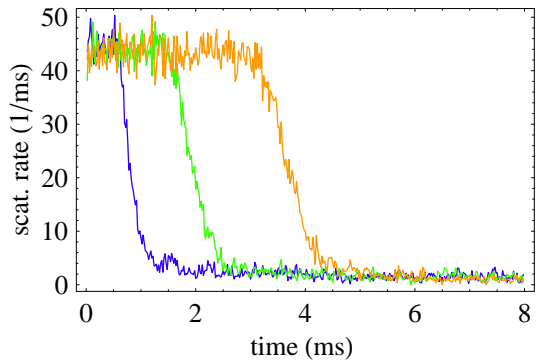}\rule{5mm}{0pt}
\caption{\label{fig:Scvst} (color online) Scattering of pump photons versus time. The lattice and pumping beam parameters are the same as in Fig.2.  The curves on the plot are results averaging 200 single atom curves. The three curves correspond to initial velocities of 1, 1.5, 2 m/s (blue, green and orange curves, respectively). }
\end{figure}

\begin{figure}[ht]
\includegraphics[width=70mm]{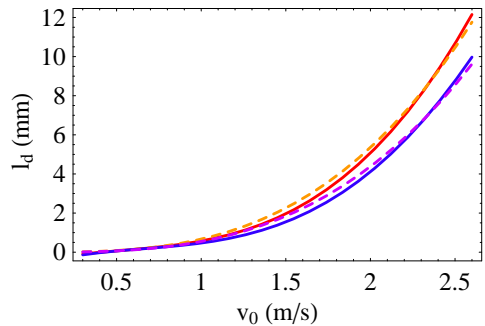}\rule{5mm}{0pt}
\caption{\label{fig:zfvsv0} (color online) Deceleration length versus initial velocity for 2 different lattice trap depths. The lattice potential is created by 20W and 30 W retro-reflected beam focused down to the 50 $\mu$m waist (corresponding to 549 and 824 $\mu$K, red and blue curves respectively). Dashed orange and purple curves are $v_{0}^{3}$ fits of the numerical data.}
\end{figure}

\begin{figure}[ht]
\includegraphics[width=70mm]{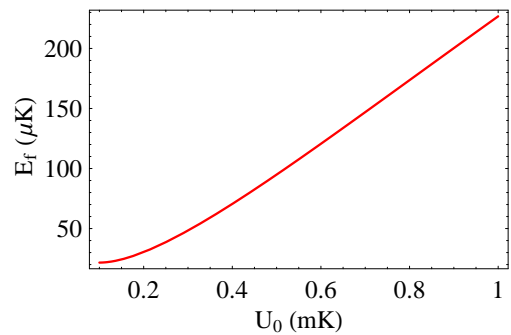}\rule{5mm}{0pt}
\caption{\label{fig:efvsu0} (color online) Final atom energies as a function of of the lattice potential trap depth. The lattice potentials is created by 20 W retro-reflected beam focused down to the 50 $\mu$m waist. Atom have been decelerated and spent trapped lattice sites of about 5 ms, after that their full energy
do not change noticeably. The full energy is a sum of kinetic and potential parts, where zero for potential energy is set to the minima of the lattice potentials.  The pumping beam is tuned near resonantly with $\delta=- \gamma/2$ and s=5.  }
\end{figure}

In Fig. 5 I plot a final atom energy as a function of the lattice potential trap depth. The full atom energy is extracted after the deceleration and equilibration in the lattice sites. 
Photon scattering stops due to the large ac Stark shift as soon as atom energy becomes noticeably smaller
than the lattice potential.  This in turn leads to termination of the cooling process.
 Although it prevents atom energies from reaching the Doppler energy (despite the different cooling mechanism this method also is Doppler-limited as shown in  \cite{Ivanov2011}), this also implies trapped atoms do not scatter any cooling photons,
thus atoms will not suffer from loss mechanisms crucial for MOTs, such as photo-assisted collisions.
The atom energy shows roughly linear dependence on the lattice trap depth except in the regime of shallow lattice depth in which case the final energies became comparable with the Doppler energy (15 $\mu$K for the  $\delta=- \gamma/2$ and s=5). However, lowering the final temperature comes with price of increased photon scattering because of reduced ac Stark shift.

\subsection{3D case}

The proposed scheme can be straightforwardly expanded to the 3 dimensional case. However, the probability to absorb a pumping photon is drastically reduced, indeed in the approximation of $\hbar \gamma \ll U_{d}$, the probability to absorb a photon scales like $(\hbar \gamma/U_{d})^{D}$, where $D$ is the dimensionality. Since the proposed method relies on the assumption that $U_{d} \gg \hbar \gamma  $ in order to ensure selective pumping, its efficiency drops in the 3D case.
One of the possible ways to circumvent this is to use a pulse scheme where optical lattice beams are switched on and off with a frequency high compared to the time scale of the atom motion. In this case the pumping efficiency decrease by a factor of 3, instead of  $(\hbar \gamma/U_{d})^{2}$.

In case of an atomic beam, i.e. a source atoms with a high brightness and low divergence, one can use a more practical simplified scheme. Deceleration in the longitudinal direction can be performed using the proposed method while collimation in transverse directions is done using 2D molasses, which simultaneously provides pumping into the excited state. Numerical simulation shows that such a method provides deceleration and cooling of the atomic beam, however atoms tend to diffuse in transverse directions on the time scale of several milliseconds. This occurs because the strong ac Stark shift prevents efficient Doppler cooling in the optical lattice and outside the optical lattice there is no force providing a transverse confinement.  In \cite{Aghajani09}, such confinement was provided by loading atoms into a magnetic guide.
It is not particularly important how the confinement is organized as long as it does not introduce a strong additional space-varying ac Stark shift. One can, for instance, picture a far-off-resonant, near-magic-wavelength beam spatially overlapped with the optical lattice. I consider the case of a blue-detuned hollow beam. Various methods for creating such beams are described in \cite{Ozeri99,Kulin01,Olson07,Isenhower09,Xu10}. The axis of such a beam is overlapped with the axis of the optical lattice with the hollow beam having a larger radius. In Fig. 6, I present the dependence of the full atom energy versus time. In order to construct a somewhat realistic experimental situation, the initial position of each atom is randomly set within the window $[-w_{0},w_{0}]$ in the $x$ and $y$ axes, i.e. the axes perpendicular to the axis of the optical lattice. In the $z$ direction, all atoms start at the position of 1 mm  with a random uncertainty of $\pm100$ $\mu$m. The atom velocities include the constant part $v_{0}$ present only in the z-direction and a random velocity spread in all directions that correspond to the a finite temperature assumed here to be 20 $\mu$K.

\begin{figure}[ht]
\includegraphics[width=74mm]{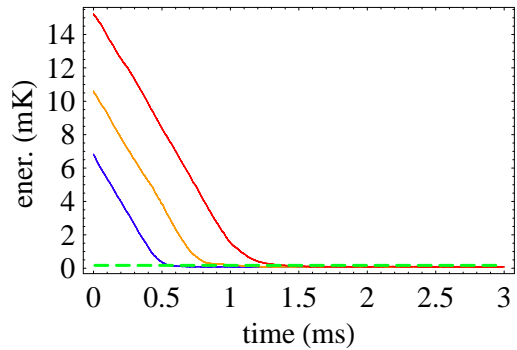}\rule{5mm}{0pt}
\caption{\label{fig:EnervsT} (color online) Energy of Yb atom versus time.  Here $\Delta t = 0.2$ ns. The lattice potential is created by 25 W retro-reflected beam focused down to a 100 $\mu$m waist at $\lambda$=1064 nm. Additionally a hollow blue detuned beam is added, with a radius of 200 $\mu$m. The pumping beam is tuned near resonance with $\delta=- \gamma/2$ and s=2. The three curves correspond to initial velocities of 0.8, 1.0, 1.2 m/s (blue, orange and red curves respectively ) directed along the optical lattice axis. }
\end{figure}

All atoms end up trapped with final energies of $k_{B}\times$75 $\mu$K as a results of this numerical experiment. This can be translated to a temperature of 25 $\mu$K ($\varepsilon=3k_{B}T $). This temperature is still an order of magnitude higher compared to the typical temperature required to reach the Lamb-Dicke regime. The scattering rate of pumping light for trapped atoms is about few photons per millisecond, i.e. at least 2 orders of magnitude smaller than in conventional MOTs.

\section{Possible Applications}

The described method can be straightforwardly used in experiment.
The number of atoms prepared in such an experiment directly depends on atomic beam parameters such as a flux, divergence \textit{e.t.c.}
An ytterbium oven in combination with differential pumping can provide a high flux, low divergence atomic beam.  In such an arrangement, atoms
have velocity spread in $z$-direction essentially defined by the oven temperature. An oven temperature of 500 $^{0}$C implies a mean thermal velocity of
$v_{m}=\sqrt{8k_{B}T/\pi m}$=305 m/s. In this case the most realistic scenario is trapping the slow portion of the Maxwell distribution.
Since only atoms passing through the optical lattice beam have only a chance to be decelerated and trapped, there is no reason to make the diameter of the oven aperture much larger than a diameter of optical lattice beams. This is advantageous for maintaining a low background vapor pressure and a low divergence of the atom beam.
At the temperature 500 $^{0}$C  the pressure of Yb vapor is $\simeq 10^{-1}$ Torr. A combination of such an oven with a set of collimation apertures can provide
good differential pumping and a well collimated, high flux atom beam with a divergence of 50 mrad or less. Details on possible ytterbium oven designs can be found for example in \cite{Loftus2001}.

Equipped with results of numerical simulations and the knowledge of the oven design I estimate the results of an experiment with concrete and realistic parameters.
For the deceleration length $l_{d}$ of 5 cm and with the parameters described in previous section the capturing velocity is about $v_{c}$=6 m/sec. In the case when the velocity $v_{c}$ is much smaller than mean thermal velocity $v_{m}$, the fraction of atoms in the Maxwell distribution having velocity below $v_{c}$ scales like $(v_{c}/v_{m})^{3}$ or more precisely equals to $\left(\frac{ 2 m v_{c}^{2}}{3 \sqrt{\pi} k_{B} T}\right)^{3/2}$.
For the given parameters the fraction of atoms that are slow enough to be decelerated and trapped is $7\times10^{-6}$.
 Overall, one expects the loading rate of about $5.6 \times 10^{7}$ atoms per second.
 For simplicity I assume the main atom loss mechanism is background gas collisions resulting in atom lifetimes of 10 sec. This limits the trapped atom number to $\sim 5.6\times10^{8}$ atoms.
 These atoms are spread over 5 cm of the deceleration length, i.e. over 10$^{4}$ lattice sites in pancake-shape samples. For the parameters used in the previous section,  an equilibrium  temperature of 25 $\mu$K.
 The estimated peak density reaches $n_{0}=\frac{N \bar{\omega}^{3}}{(2 \pi k_{B} T/m)^{3/2}}=7.4\times10^{12}$ $cm^{-3}$, with a phase-space density of $\rho_{PSD}=n_{0}\lambda_{dB}^{3} \simeq 1.5 \times 10^{-4}$.
 Loading rate of $5.6 \times 10^{7}$ s$^{-1}$ is about a factor of 100 higher that demonstrated in  \cite{Hansen2011} and roughly a factor of 20 higher than reported in  \cite{Takasu2003}.
Although similar temperatures are  observed in conventional compressed Yb MOTs, the presented method promises higher densities. Further, a conventional Yb MOT operating on
narrow 556 nm transition typically requires a Zeeman slower.
These conditions also appear to be favorable for further evaporative cooling with a collision rate on the order of 150 collisions per second. At such densities 3-body losses are still not substantial \cite{Takasu2003}. To avoid undesired losses caused by photo-associated collisions the atoms have to be removed from the optical pumping region by introducing a frequency shift between lattice beams, i.e. performing an optical elevator.


\section{Conclusions}

 I have described a dissipative deceleration of atoms  moving through an optical lattice potential that exploits the lattice-induced differential ac Stark shift of two levels coupled through an optical transition. The presented cooling scheme resembles other Sisyphus cooling  methods, however it combines a deceleration and continuous loading of the atom beam into the conservative potential of optical lattices. I use the example of an $^{174}$Yb atom beam loaded into the 1064 nm  optical lattice. In transverse directions atoms interact with near resonant 556 nm beams which pumps the atoms into the excited state while cooling in the transverse directions.
Temperatures as low as $\sim$25 $\mu$K are reachable in time scales of few milliseconds. The prepared atom sample can be outcoupled using the optical elevator arrangement.
The present scheme can be straightforwardly applied to other atomic species.
 The electronic transitions with linewith $~$ 100kHz are available for various atomic species.
 In alkali atoms, the usual D2 line $nS_{1/2}\longrightarrow nP_{3/2}$ is rather broad, but the narrower $nS_{1/2}\longrightarrow (n+1)P_{3/2}$ line may be utilized. For example laser cooling of lithium and potassium was recently demonstrated using such transitions \cite{Duarte2011,McKay2011} (the linewidth are $\gamma=2 \pi \times 754$ kHz for lithium and  $\gamma=2 \pi \times 1.19 $ MHz for potassium).
In general, detailed studies of ac Stark shift are required for the levels coupled by the pumping transition, however in the vicinity of the pumping transition, the ac Stark shifts are dominated by this transition and at the red side of it ac Stark shifts have the right configuration for use of the proposed method.
Strontium atoms, which posses the narrow  $^{1}S_0 \rightarrow {^{3}P_1}$ transition at $\lambda_{\rm Sr}=689$nm with linewidth $\gamma_{\rm Sr}=2 \pi \times 7.4$ kHz, are another example of atomic specie applicable for the presented scheme.
In case of Sr, the optical lattice wavelengths in the range from  800nm to 2 $\mu$m meet the criteria for the ac Stark shifts  \cite{Katori1999,Ivanov2011}.

 One of the important advantages of this method is a low number of photons that the atoms need to scatter in order to achieve substantial cooling. During simulations it was found that about 100 photons are sufficient to decelerate and trap atoms with 1 m/s initial velocity. This softens requirements for the branching ratio of the cooling transition and it makes this method attractive for
cooling species that do not possess near-cycling transitions with a high branching ratio. An appealing prospect is cooling of heteronuclear molecules. Recent work in the DeMille group has demonstrated laser cooling of heteronuclear molecules \cite{Shuman2010} on a transition with a moderately high branching ratio adequate for cooling using the proposed method.

I thank Kara Maller and Alex Khramov for a critical reading of the manuscript, and Subhadeep Gupta for
useful discussions.

\end{document}